    \documentclass[%
    aip,
    amsmath,amssymb,
    reprint,%
    ]{revtex4-1}

    \usepackage{graphicx}% Include figure files
    \usepackage{dcolumn}% Align table columns on decimal point
    \usepackage{bm}% bold math
    \usepackage[utf8]{inputenc}
    \usepackage[T1]{fontenc}
    \usepackage{mathptmx}
    \usepackage{etoolbox}

    \usepackage{subcaption} 
    \usepackage[linesnumbered, ruled, vlined]{algorithm2e}
    \usepackage[colorlinks=true,linkcolor=blue,citecolor=blue,urlcolor=blue]{hyperref}
    \usepackage[dvipsnames]{xcolor}
    \usepackage{caption}
    \makeatletter
    \def\@email#1#2{%
    \endgroup
    \patchcmd{\titleblock@produce}
    {\frontmatter@RRAPformat}
    {\frontmatter@RRAPformat{\produce@RRAP{*#1\href{mailto:#2}{#2}}}\frontmatter@RRAPformat}
    {}{}
    }%
    \makeatother

\begin{document}

    \preprint{AIP/123-QED}

    \title[AeroDiT]{AeroDiT: Diffusion Transformers for Reynolds-Averaged Navier-Stokes Simulations of Airfoil Flows}
    % Force line breaks with \\

    \author{Chunyang Wang}
    \altaffiliation{These authors contributed equally to this work.}
    \affiliation{Science42 Technology, Beijing, China}
    \affiliation{Imperial College London, London, UK}

    \author{Biyue Pan}
    \altaffiliation{These authors contributed equally to this work.}
    \affiliation{Science42 Technology, Beijing, China}
    
    \author{Zhibo Dai}
    \affiliation{Science42 Technology, Beijing, China}

    \author{Yudi Cai}
    \affiliation{Science42 Technology, Beijing, China}

    \author{Yuhao Ma}
    \affiliation{Science42 Technology, Beijing, China}

    \author{Hao Zheng}
    \affiliation{Science42 Technology, Beijing, China}

    \author{Dixia Fan}
    \email{fandixia@westlake.edu.cn}
    \thanks{Corresponding author}
    \affiliation{School of Engineering, Westlake University, Hangzhou, China}
    \affiliation{Institute of Advanced Technology, Westlake Institute for Advanced Study, Hangzhou, China}

    \author{Hui Xiang}
    \email{xianghui@science42.tech}
    \thanks{Corresponding author}
    \affiliation{Science42 Technology, Beijing, China}

    \date{\today}% It is always \today, today,

    \begin{abstract}

    Real-time and accurate prediction of aerodynamic flow fields around airfoils is crucial for flow control and aerodynamic optimization. However, achieving this remains challenging due to the high computational costs and the non-linear nature of flow physics. Traditional Computational Fluid Dynamics (CFD) methods face limitations in balancing computational efficiency and accuracy, hindering their application in real-time scenarios. To address these challenges, this study presents AeroDiT, a novel surrogate model that integrates scalable diffusion models with transformer architectures to address these challenges. Trained on Reynolds-Averaged Navier-Stokes (RANS) simulation data for high Reynolds-number airfoil flows, AeroDiT accurately captures complex flow patterns while enabling real-time predictions. The model demonstrates impressive performance, with mean relative $L_2$ errors of 0.1, 0.025, and 0.050 for pressure \(p\) and velocity components \(u_x, u_y\), confirming its reliability. To further enhance physical consistency, we incorporate explicit physics-informed losses based on RANS residuals, including mass and momentum conservation constraints. The transformer-based structure allows for real-time predictions within seconds, enabling efficient aerodynamic simulations. This work underscores the potential of generative machine learning techniques to advance computational fluid dynamics, offering potential solutions to challenges in simulating high-fidelity aerodynamic flows.

    \end{abstract}

    \maketitle

    \section{Introduction}\label{sec1}
    %%background
    Computational Fluid Dynamics (CFD) plays a pivotal role in solving and analyzing fluid flow problems, serving as an indispensable tool across diverse fields such as aerospace engineering, environmental science, industrial design, and biomedical research. Despite its broad applicability, a major challenge of CFD lies in its high computational cost. Techniques like Direct Numerical Simulation (DNS) and Large Eddy Simulation (LES) are computationally intensive, often making them impractical for many applications. While Reynolds-Averaged Navier-Stokes (RANS) models offer a more efficient alternative, they still fall short for real-time applications. These constraints limit CFD's applicability in areas such as optimal flow control and airfoil optimization. To overcome these limitations, there is a critical need for the development of alternative methods and frameworks that improve computational efficiency without compromising simulation accuracy.

    %%Research problem
    In recent years, the emergence of deep learning techniques has opened new avenues for efficient flow field predictions. A growing body of research has integrated deep learning methods with fluid flow simulation, yielding promising results. Notably, neural network architectures such as U-net \cite{guo2016convolutional, sekar2019fast, ribeiro2020deepcfd, chen2021numerical, cai2022fast}, generative adversarial networks (GANs) \cite{goodfellow2020generative, chen2020airfoil, haizhou2022generative}, and physics-informed neural networks (PINNs) \cite{raissi2019physics, cai2021physics, chen2021physics, zhang2023physics} have shown considerable potential in tasks such as flow field reconstruction. These approaches demonstrate the power of deep learning to capture the complex dynamics of fluid flow, signaling a shift in the future of CFD research.  Guo et al.\cite{guo2016convolutional} proposed a convolutional neural network (CNN)-based surrogate model for real-time prediction of steady laminar flow fields, achieving significantly faster velocity field estimation compared to traditional CFD solvers while maintaining low error rates. Wu et al.\cite{haizhou2022generative} proposed a data-augmented Generative Adversarial Network (daGAN) for rapid and accurate airfoil flow field prediction, demonstrating strong generalization capabilities with sparse training data through pre-training and fine-tuning modules. Raissi et al.\cite{raissi2019physics} introduced physics-informed neural networks , a deep learning framework that integrates physical laws governed by nonlinear partial differential equations to solve forward and inverse problems, enabling data-efficient spatio-temporal function approximation and scientific discovery even in small data regimes. 

    The deep learning landscape is still rapidly evolving, with newer architectures such as Transformers \cite{vaswani2017attention, khan2022transformers, lin2022survey} and denoising diffusion probabilistic models (DDPMs) \cite{ho2020denoising, yang2023denoising, yang2023denoising}-along with scalable diffusion models integrated with Transformers (DiT) \cite{peebles2023scalable}, showing superior performance in tasks like image generation. This evolution raises an important question: \textit{Can the latest advancements in machine learning enable real-time, accurate predictions of flow fields, effectively addressing the challenges of computational efficiency and accuracy?}

    Although recent advancements using DDPMs have shown promising results in fluid dynamics prediction, these models have primarily been applied to relatively simple flow fields. For instance, methods like FluidDiff \cite{yang2023denoising} and the approach by Shu et al. \cite{shu2023physics} have demonstrated success in prediction and reconstruction tasks, but they are generally limited to simple flow conditions. Furthermore, the time-consuming sampling process inherent in U-Net-based DDPMs makes them unsuitable for real-time predictions. These limitations highlight the urgent need for more efficient and practical methods capable of handling complex, real-time fluid field predictions.

    Efficient and accurate prediction of high Reynolds number airflow around airfoils is crucial for advancements in aerodynamics research and the aviation industry, as it directly impacts airfoil optimization, optimal flow control, and the design of more efficient aerodynamic systems. The application of deep learning in aerodynamics has evolved progressively from foundational flow field prediction to efficient optimization design. Early research \cite{sekar2019fast} established a speed foundation for replacing traditional CFD through a hybrid CNN-MLP architecture (where CNNs extract geometric features and MLPs map flow parameters), enabling real-time flow field prediction, though limited to steady laminar flows. To address data dependency issues, Wu et al.\cite{haizhou2022generative} introduced the daGAN stucture, leveraging a two-phase training mechanism (cGAN pre-training followed by dual-generator fine-tuning) to overcome sparse data constraints and enhance generalization under small-sample conditions. Li et al. \cite{kaandorp2020data} developed Deep Learning-based Geometric Filtering (DLGF), utilizing autoencoder-based compression of the design space to reduce variable dimensionality from hundreds to tens, thereby circumventing the curse of dimensionality and accelerating optimization workflows. Within this framework, Wang et al. \cite{muchen2022airfoil} integrated a CNN-based Reduced Order Model (ROM) with a fusion strategy of pooling dimensionality reduction and radial basis functions to handle nonlinear flow optimization under large disturbances. Concurrently, Li et al. \cite{li2022low} addressed low-Reynolds-number separation bubbles by integrating the XFOIL solver with a transition-sensitive turbulence model, achieving efficient airfoil design for UAVs through tailored modal decomposition.

    While these works mentioned above collectively aim to replace high-cost simulation iterations via a paradigm integrating physical constraints with data-driven intelligence, they exhibit limitations: conventional generative models (e.g., GANs, DDPMs) inadequately capture complex separated flows in high-Reynolds-number turbulence; U-Net based diffusion models suffer from inefficient iterative sampling that precludes real-time application; and purely data-driven approaches show weak extrapolation capability under extreme Reynolds numbers while lacking embedded physical constraints, leading to localized errors near critical airfoil surfaces. However, due to the complexity of the flow-characterized by turbulence and flow separation-achieving both efficiency and accuracy in these predictions remains a significant challenge.

    This study introduces a novel methodology for aerodynamic flow-field simulation Adaptive Diffusion Transformers for Airfoil Flow Simulation (AeroDiT) which employs diffusion transformers to learn from high-fidelity Reynolds-Averaged Navier-Stokes (RANS) simulation data. By treating the complex flow-field distributions as high-dimensional probability spaces and leveraging a generative approach, AeroDiT aims to dramatically reduce computational costs while preserving predictive accuracy at high Reynolds numbers. Compared to conventional computational fluid dynamics (CFD) techniques, AeroDiT's innovations extend both the theoretical and practical frontiers of aerodynamic simulation, yielding the following key contributions:

    \textbf{1. Pioneering Application of Diffusion Transformers in Aerodynamic Flow Prediction.} This work represents the first utilization of Scalable Diffusion Models with Transformers (DiTs) as surrogate models for predicting aerodynamic flow fields. Traditional CFD approaches to high-Reynolds-number turbulence require fine spatial discretization and iterative solvers, resulting in prohibitive computational overhead. AeroDiT reframes the flow-field as a high-dimensional latent distribution and embeds multiscale turbulent structures into a diffusion-based generative framework. The transformer's self-attention mechanism captures interactions across all spatial scales, enabling efficient reconstruction of the entire flow field from learned statistical representations. From a theoretical standpoint, this extends the application paradigm of generative models within physical sciences and demonstrates a data-driven route to approximately solve partial differential equations (PDEs), thereby circumventing the heavy numerical discretization typical of classical methods.

    \textbf{2. Experimentally Validated, High-Fidelity, Real-Time Prediction Capability.} During validation, AeroDiT exhibits exceptional accuracy and real-time performance. Across a variety of angles of attack, freestream velocities, and representative airfoil geometries, the model consistently achieves relative L2 prediction errors on velocity fields, pressure distributions, and turbulence metrics (such as vorticity) on the order of only a few percent. Simultaneously, the generative inference process can reconstruct a complete flow field in a matter of seconds-many orders of magnitude faster than RANS-based CFD. Consequently, AeroDiT holds immense promise for applications requiring massive parametric sweeps, real-time control, or rapid design iteration. It can accelerate the design optimization cycle in aerospace engineering and furnish reliable surrogate models for data-driven, multidisciplinary design optimization.

    \textbf{3. Integration of Physical Priors to Ensure Consistency with Governing Equations.} Beyond empirical accuracy and speed, AeroDiT's core innovation lies in blending generative modeling with physics-informed constraints. By embedding RANS-derived physical priors during training, the network continually corrects deviations from conservation laws, preventing unbounded extrapolation errors in out-of-distribution regimes. Such a methodology paves the way for extending generative deep learning to more complex flow problems (e.g., multiphase flows, combustion, three-dimensional wing arrays), offering a general framework for achieving a balance between physical fidelity and statistical expressivity.

    The remainder of the paper is organized as follows: Section~\ref{section:Related Work} reviews related work our work has been built on, and provides an overview of the background information. Section~\ref{Section: Methodology} outlines the fundamental methodology used in this work. Section~\ref{Section:Experimental} presents the experimental evaluation, detailing datasets, configurations, and results obtained through extensive testing. Section~\ref{Section:Discussion} discusses the implications of the results, highlighting the strengths and limitations of the proposed approach while suggesting future research directions.

    \section{Related Work}\label{section:Related Work}
    \subsection{Diffusion Models in Fluid Dynamics}
    Recent advances in deep learning-based fluid dynamics prediction have utilized generative models to address the challenges of simulating complex flow fields. Yang et al. \cite{yang2023denoising} introduce FluidDiff, a novel generative model based on Denoising Diffusion Probabilistic Models (DDPMs). FluidDiff employs a diffusion process to learn a high-dimensional representation of dynamic systems and uses Langevin sampling to predict flow states under specific initial conditions. Trained on finite, discrete fluid simulation data, FluidDiff achieves competitive performance, accurately predicting test data without the need for explicit encoding of the underlying physical system' prior knowledge. Similarly, Shu et al. \cite{shu2023physics} propose a DDPM-based model for flow field super-resolution, trained on high-fidelity data to reconstruct detailed flow fields from low-fidelity or sparsely sampled data. This model also incorporates physics-informed conditioning through partial differential equations (PDEs), improving its predictive accuracy. Hu et al. \cite{hu2024generative} introduce the Geometry-to-Flow (G2F) diffusion model, which generates flow fields around obstacles by conditioning on obstacle geometry. Zhou et al. \cite{zhou2024text2pde} present Text2PDE, a latent diffusion model framework for PDE simulation that uses autoencoders and text-based conditioning to generate full spatio-temporal solutions. Liu et al.\cite{liu2024uncertainty} proposed an uncertainty-aware surrogate model for turbulence simulations using DDPMs, demonstrating superior accuracy and uncertainty quantification compared to Bayesian neural networks and heteroscedastic models. While these studies show promising results, many of them primarily focus on simpler flow fields, leaving room for further exploration in more complex and high-dimensional flow prediction scenarios. To address these limitations, there is increasing interest in developing more advanced architectures capable of modeling such complexities, such as the Diffusion Transformer, which is introduced in the next section.

    \subsection{Diffusion Transformer}
    In recent years, Transformers have achieved remarkable success across various domains, including natural language processing and computer vision \cite{vaswani2017attention}. Meanwhile, diffusion models, as a class of generative models, have shown significant potential in image generation tasks. The Diffusion Transformer (DiT) combines the strengths of Transformers with a diffusion-based architecture, demonstrating outstanding performance in diverse applications \cite{peebles2023scalable}. Despite these advances, the application of DiT models to aerodynamic simulation remains unexplored. To the best of our knowledge, this study represents the first application of a DiT-based model trained on Reynolds-averaged Navier-Stokes (RANS) simulation data, with a particular focus on predicting flow around airfoils. In addition, in line with prior DiT-based work \cite{ovadia2023real}, our study addresses the challenge of real-time inference. A detailed description of the methodology is provided in the following section.

    \section{Methodology} \label{Section: Methodology}

    \begin{figure*}[!t]
        \centering
        \includegraphics[width=0.95\linewidth]{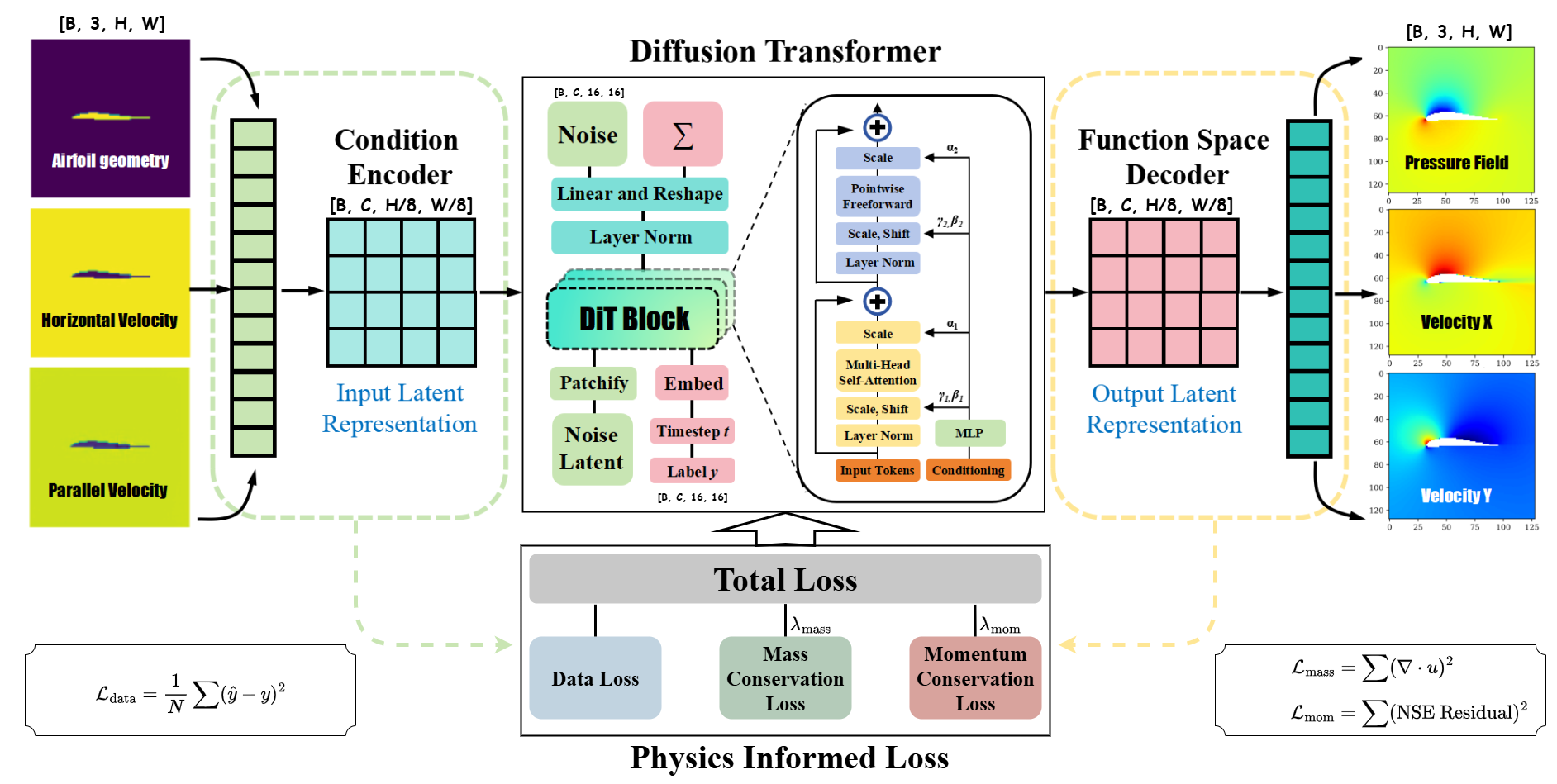}
        \caption{Overview of the AeroDiT framework. The input consists of three channels: airfoil geometry mask, horizontal velocity ($u_x$), and vertical velocity ($u_y$), concatenated as a tensor of shape $[B, 3, H, W]$. These are processed by the \textbf{Condition Encoder}, which downsamples the input into a latent representation. Simultaneously, the ground truth pressure and velocity fields are encoded into a target latent space by a VAE. The \textbf{Diffusion Transformer (DiT)} operates on this latent space by iteratively denoising the noisy latent using transformer blocks conditioned on the encoded input and timestep embeddings. The output latent is decoded by the \textbf{Function Space Decoder} to reconstruct the pressure and velocity fields. The model is optimized using a combination of data loss, mass conservation loss, and momentum conservation loss to ensure physical consistency.}
        \label{fig:main}
    \end{figure*}

    This study employs Diffusion Transformers to predict the airfoil flow field at high Reynolds numbers. The framework of DiTs is shown in Figure \ref{fig:main}. By leveraging the generative nature of diffusion models, DiTs serve as a surrogate model that learns the statistical distribution of flow field data. Once trained, the model can predict velocity and pressure fields based on input airfoil geometry and flow conditions.

    The training process consists of two key steps: forward diffusion and reverse denoising, as shown in Figure \ref{fig:main}. In the forward diffusion step, Gaussian noise is progressively added to the input flow field, gradually transforming it into a noise distribution. The reverse denoising process then reconstructs the original flow field iteratively, utilizing Bayesian inference to model the transition probabilities. Together, these processes enable the model to effectively approximate the underlying distribution and generate accurate predictions. 

    We then introduce the notations, model architecture, diffusion processes, loss functions, and training procedure employed in AeroDiT.

    \subsection{Notations}\label{sec:notatio}

    To ensure clarity, we define the key symbols used throughout this study:
    \begin{itemize}
        \item \(\boldsymbol{x}\): Flow field, including pressure \(\mathbf{p}\) and velocity components \(u_x\) (along \(x\)-axis) and \(u_y\) (along \(y\)-axis).
        \item \(\boldsymbol{\Psi}\): Input conditions, including airfoil geometry and inflow conditions \(u_x\), \(u_y\).
        %\item \(\mathbf{u}\): Velocity vector, with components \(u_x\) and \(u_y\).
        \item \(\mathbf{p}\): Pressure field. Here, we bold \(\mathbf{p}\) to distinguish it from the probability distribution $p$.
        %\item \(\rho\): Fluid density.
        %\item \(\nu\): Kinematic viscosity.
        %\item \(\boldsymbol{\tau}\): Viscous stress tensor.
        %\item \(\mathbf{f}\): External forces acting on the flow field.
        %\item \(B(\mathbf{u}, \mathbf{p}), \rho\): Boundary conditions applied to the computational domain boundary \(\partial \Omega\).
        \item \(\mathbf{S}(\boldsymbol{\Psi})\): The AeroDiT surrogate model.
        \item \(\mathbf{R}(\boldsymbol{\Psi})\): The real physical system.
        \item \(\epsilon \sim \mathcal{N}(0, \mathbf{I})\): Gaussian noise added during forward diffusion.
        \item \(\alpha_t, \bar{\alpha}_t\): Variance scaling factors in the diffusion process.
    \end{itemize}

    \subsection{Learning Target}\label{sec:Learning Target}
    The primary objective is to develop a surrogate model, \(\mathbf{S}\), that approximates the real physical system, \(\mathbf{R}\), governed by the Navier-Stokes equations.

    As discussed before, solving these equations numerically for real flow fields $\mathbf{R}(\boldsymbol{\Psi})$ is computationally expensive. To address this, the learning target is defined as identifying the optimal parameters \(\theta\) of the surrogate model \(\mathbf{S}\) such that:
    \begin{equation}
    \boldsymbol{x} = \mathbf{S}_{\theta}(\boldsymbol{\Psi}) \approx \mathbf{R}(\boldsymbol{\Psi}),
    \end{equation}

    From the perspective of generative modelling, this can be achieved by learning the conditional distribution \(p(\boldsymbol{x} | \boldsymbol{\Psi})\) from our dataset $\boldsymbol{D}$, which captures the statistical relationship between the input conditions $\boldsymbol{\Psi}$ and the resulting flow field $\boldsymbol{x}$. To approximate the real physical system, the model optimizes parameters $\theta$ by maximizing the likelihood of observed flowfield data:
    \begin{equation}
        \max_\theta \mathbb{E}_{\boldsymbol{x}, \boldsymbol{\Psi}} \big[\log p_\theta(\boldsymbol{x} | \boldsymbol{\Psi})\big].
    \end{equation}

    Therefore we formulate our learning target as learning the conditional distribution \(p(\boldsymbol{x} | \boldsymbol{\Psi})\) from our dataset $\boldsymbol{D}$. By learning this distribution, the model can generate accurate predictions efficiently, bypassing the computational complexity of solving the Navier-Stokes equations numerically. To approximate such distribution, our AeroDiT model goes through a forward diffusion process and a reverse denoising process, as discussed in the following sections.

    \begin{figure}
        \centering
        \includegraphics[width=1\linewidth]{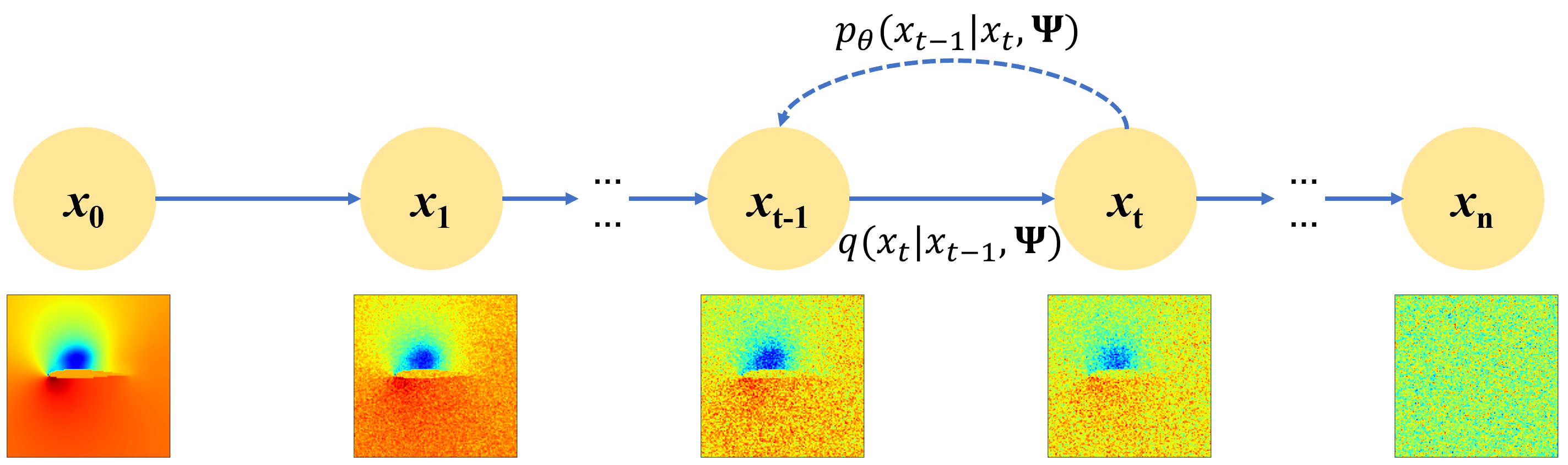}
        \caption{Illustration of the forward and backward diffusion process.}
        \label{fig:1}
    \end{figure}

    \subsection{Model Architecture} \label{sec:architecture}

    Figure~\ref{fig:main} illustrates the overall architecture of AeroDiT, which consists of three main modules: a condition encoder, a transformer-based diffusion module (DiT), and a function space decoder. The model operates entirely in the latent space, enabling compact yet expressive representations of high-resolution flow fields.

\begin{figure}[htbp]
        \centering
        \includegraphics[width=0.4\textwidth]{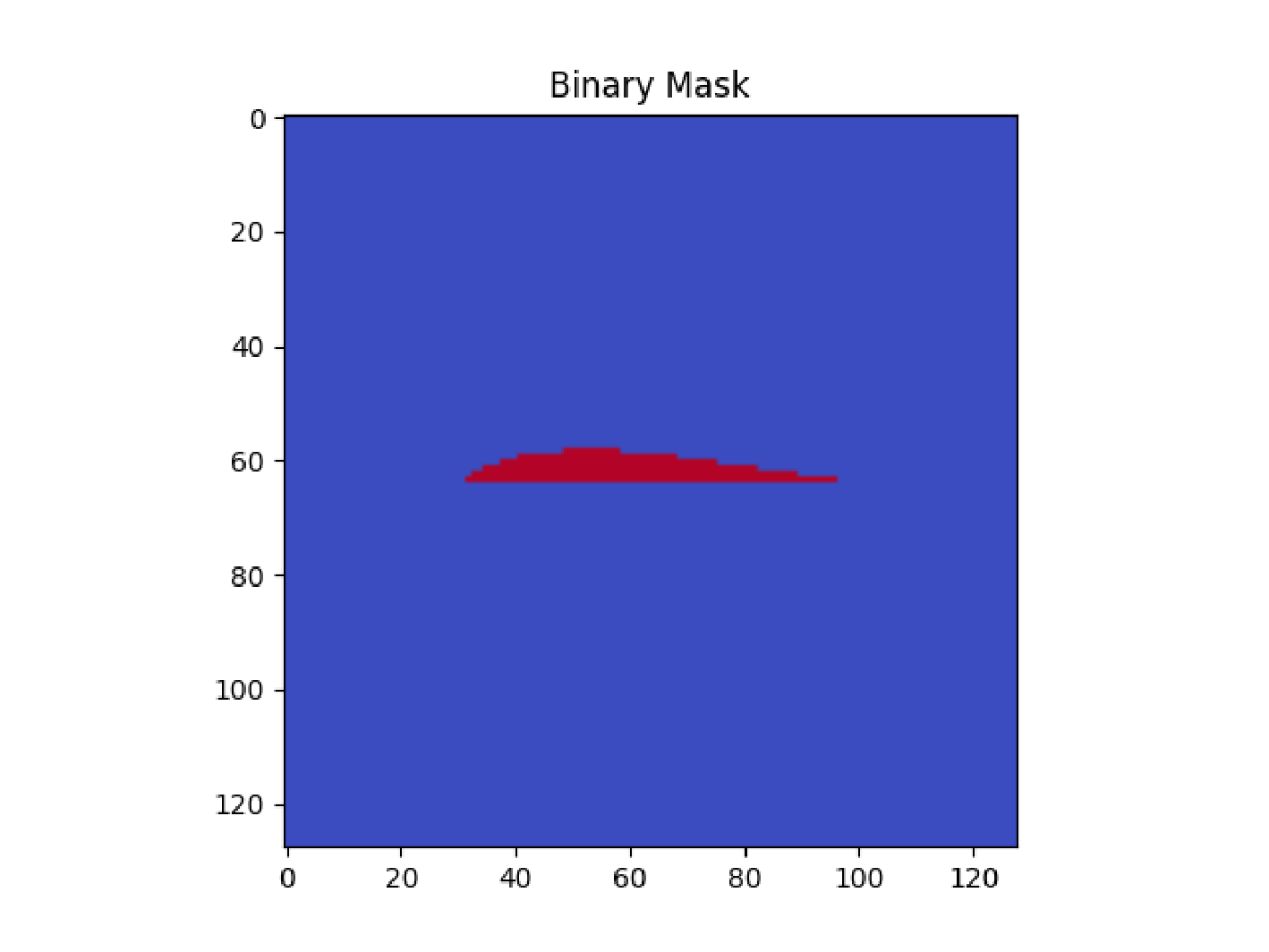}
        \caption{Wing geometry. (The red part shows the wing and the blue part shows the flow field)}
        \label{fig:wing_geometry}
    \end{figure}

    \textbf{Input tensor structure.}  
    The input to the model is a tensor of shape $[B, 3, H, W]$, where $B$ is the batch size, and $H \times W$ denotes the spatial resolution of the flow field (typically $128 \times 128$). The three input channels correspond to:  
    (1) the airfoil geometry mask (binary-valued), as illustrated in Figure~\ref{fig:wing_geometry},
    (2) inflow velocity in the $x$-direction ($u_x$), and  
    (3) inflow velocity in the $y$-direction ($u_y$).

    Instead of directly feeding raw spatial inputs into the diffusion model, we first encode them using a dedicated \textit{condition encoder}, composed of four convolutional layers with stride 2 and ReLU activations. This module downscales the input to a compact latent condition embedding of shape $[B, C, H', W']$, where $C$ is the latent channel dimension and $H', W' = H/8, W/8$. This design reduces memory overhead and enhances generalization by extracting task-relevant features while suppressing irrelevant artifacts. 

    \textbf{Target VAE encoder.}  
    In parallel, the ground truth outputs-pressure $p$, and velocity components $u_x$, $u_y$-are passed through a variational autoencoder (VAE) encoder of identical structure to obtain a latent representation of the same shape $[B, C, H', W']$. The total number of parameters in the encoder-decoder pair is approximately 1 million, offering a compact and efficient representation of high-fidelity simulation data. While the encoder is trained jointly with the diffusion model to extract meaningful conditional features, the decoder is pre-trained as part of the VAE and kept frozen during DiT training, forming an asymmetric training strategy that stabilizes the reconstruction process and improves convergence.

    \textbf{Latent-space diffusion via DiT.}  
    The core of AeroDiT is a transformer-based diffusion model (DiT) that iteratively denoises the latent representation. At each reverse timestep, the DiT takes three inputs:  
    (1) the current noisy latent,  
    (2) the condition embedding from the encoder, and  
    (3) a learned timestep embedding.

    We adopt a lightweight DiT variant consisting of 6 transformer blocks, each with 8 attention heads and feedforward sublayers. Self-attention is applied over the flattened $H' \times W'$ latent grid, allowing the model to capture both local structures and long-range dependencies-critical for accurately modeling multiscale turbulence and boundary-layer dynamics. By operating in the compressed latent space, the model achieves strong predictive accuracy while remaining computationally tractable. Compared to pretrained vision models (e.g., ViT or CLIP), which are optimized for natural images, our task-specific encoder provides physics-aware inductive bias and is trained jointly with the diffusion model to ensure consistency across modules.

    \textbf{Decoding and prediction.}  
    After iterative denoising in latent space, the resulting tensor is passed through a symmetric convolutional decoder to reconstruct the physical fields at the original resolution $[B, 3, H, W]$, corresponding to the predicted pressure and velocity fields.

    Together, this architecture enables conditional generation of high-resolution flow fields from geometry and inflow conditions, while ensuring multiscale awareness and physical consistency. The training strategy and hybrid loss functions are detailed in Section~\ref{sec:loss}.

    \subsection{Forward Diffusion Process}\label{sec:forward}
    In the forward diffusion process, we followed the canonical procedures described in literature \cite{ho2020denoising, sohl2015deep}. As shown in Figure \ref{fig:1}, we gradually add noise to the data \( \boldsymbol{x}_0 \) (the true flow field) over several time steps given the input conditions $\boldsymbol{\Psi}$, transforming it into a noisy sample \( \boldsymbol{x}_T \). The noise schedule \( \beta_t \) is pre-defined and fixed, increasing with each timestep. At each timestep \( t \), the conditional probability for adding noise is given by:

    \begin{align}
    q(\boldsymbol{x}_t | \boldsymbol{x}_{t-1}, \Psi) &= \mathcal{N}(\boldsymbol{x}_t; \sqrt{1 - \beta_t} \boldsymbol{x}_{t-1}, \beta_t \mathbf{I})
    \end{align}
    where \( \boldsymbol{x}_t \) represents the flow field at timestep \( t \), and \( \beta_t \) is the pre-defined noise variance at that timestep. The noise is added progressively until, at the final timestep \( T \), the data becomes almost entirely noise. 

    The main goal of this process is to transform the data into a noise distribution that can be efficiently learned by the reverse denoising process. Since the variable $\boldsymbol{\Psi}$ does not break the Markovian nature of the forward diffusion process, the forward diffusion process for \( t \in [1, T] \) is formally described as:

    \begin{equation}
    q(\boldsymbol{x}_t | \boldsymbol{x}_0, \Psi) = \mathcal{N}(\boldsymbol{x}_t; \sqrt{\bar{\alpha}_t} \boldsymbol{x}_0, (1 - \bar{\alpha}_t) \mathbf{I})
    \end{equation}
    where \( \bar{\alpha}_t = \prod_{i=1}^{t} (1 - \beta_i) \) and \( \boldsymbol{x}_0 \) is the original clean data. The term \( \bar{\alpha}_t \) is the cumulative product of the noise schedule, determining the level of corruption at each timestep. The noise becomes larger as \( t \) increases, leading to a greater amount of information lost by timestep \( T \). The denoising process will learn to remove the noise and hence learn the conditional distribution, as discussed in the next section.

    \subsection{Reverse Denoising Process and the Loss}\label{sec:denoising}
    In the reverse denoising process, we start with the noisy data \( \boldsymbol{x}_T \) obtained from the forward diffusion process and iteratively remove the noise to recover the original data \( \boldsymbol{x}_0 \). The ultimate goal for the denoising process is to help achieve our learning target discussed in Section \ref{sec:Learning Target}. 

    By constructing the forward diffusion process, we can employ variational inference to show that minimizing the Kullback-Leibler (KL) divergence 
    \begin{equation}
    KL\big(q(\boldsymbol{x}_{t-1} \mid \boldsymbol{x}_t, \boldsymbol{\Psi}) \,\|\, p_\theta(\boldsymbol{x}_{t-1} \mid \boldsymbol{x}_t, \boldsymbol{\Psi})\big)
    \end{equation}
    effectively maximizes the Evidence Lower Bound (ELBO) of 
    \begin{equation}
    \mathbb{E}_{\boldsymbol{x}, \boldsymbol{\Psi}} \big[\log p_\theta(\boldsymbol{x} | \boldsymbol{\Psi})\big].
    \end{equation}
    Here, the KL divergence measures how closely two probability distributions align. This insight allows us to reformulate the learning target: instead of directly maximizing 
    $\mathbb{E}_{\boldsymbol{x}, \boldsymbol{\Psi}} \big[\log p_\theta(\boldsymbol{x} | \boldsymbol{\Psi})\big]$ 
    we focus on learning the optimal parameters \( \theta \) so that \( p_\theta(\boldsymbol{x}_{t-1} | \boldsymbol{x}_t, \boldsymbol{\Psi}) \) accurately approximates \( q(\boldsymbol{x}_{t-1} | \boldsymbol{x}_t, \boldsymbol{\Psi}) \). In other words, by correctly learning the denoising process, we approximate the conditional distribution in a stepwise manner. 

    The reverse denoising process \( q(\boldsymbol{x}_{t-1} | \boldsymbol{x}_t, \boldsymbol{\Psi}) \) follows a Gaussian distribution, with its mean and variance derived as\cite{ho2020denoising}:
    \begin{align}
        \tilde{\mu}_t &= \frac{1}{\sqrt{\alpha_t}} \left( \boldsymbol{x}_t - \frac{1 - \alpha_t}{\sqrt{1 - \bar{\alpha}_t}} \boldsymbol{\epsilon} \right)\\
        \tilde{\beta}_t &= \frac{1 - \bar{\alpha}_{t-1}}{1 - \bar{\alpha}_t} \beta_t
    \end{align}

    We notice that $\tilde{\beta}_t$ depends only on the predetermined hyperparameters, and we assign such a constant value directly to the variance of \( p_\theta(\boldsymbol{x}_{t-1} | \boldsymbol{x}_t, \boldsymbol{\Psi}) \). This leaves us only with the need to adjust $\theta$ so that the mean converges. By observation, one can realize that making $\mu_{\theta}$ converge to $\tilde{\mu}$
    is equivalent to making the noise prediction converge to the noise added in the diffusion process. Therefore, from our initial learning target, we formulate our final learning loss:
    \begin{equation}
    \mathcal{L}_{MSE}(\theta) = \mathbb{E}_{t, \boldsymbol{x}_0, \boldsymbol{\Psi}, \epsilon} \left[ \| \epsilon - \epsilon_\theta(\boldsymbol{x}_t, t, \boldsymbol{\Psi}) \|^2 \right]
    \end{equation}

    In the DiT framework, the noise term, \( \epsilon_\theta(\boldsymbol{x}_t, t, \boldsymbol{\Psi}) \), is trained and predicted using a Transformer-based architecture, enabling the model to capture long-range dependencies and contextual information. This contrasts with DDPMs, which typically employ simpler architectures, such as MLPs or CNNs. By leveraging self-attention, DiT incorporates richer contextual information, improving its ability to capture dependencies and correlations within the data. This enhances the model's ability to handle complex inputs, such as structured data or sequential dependencies (e.g., time series or flow fields). Rather than directly predicting the noise, DiT predicts a denoised version of the flow field \( x_0 \), utilizing attention mechanisms. Although the predicted noise \( \epsilon_{\theta} \) is still required, it is inferred in a manner distinct from that in DDPM-based models.

    By accurately predicting the noise, the model learns the stepwise transformations required to approximate the conditional distribution $p(\boldsymbol{x} \mid \boldsymbol{\Psi})$. Based on this theoretical foundation, the following section focuses on the practical implementation of these concepts.

    \subsection{Physics Informed Loss} \label{sec:loss}

    The training of AeroDiT is guided by a hybrid loss function that combines data-driven supervision with physics-based constraints. This design enables the model to learn not only from statistical correlations in the training data, but also from the governing laws of fluid mechanics.

    The core data-driven term is the mean squared error (MSE) derived in the previous section, to enhance the physical plausibility of the generated flow fields, we incorporate additional loss terms derived from the steady-state Reynolds-Averaged Navier-Stokes (RANS) equations. For incompressible flows, the RANS equations are fundamentally composed of two conservation laws: \textbf{mass conservation} and \textbf{momentum conservation}. These two constraints fully characterize the behavior of the fluid system and are thus used as the basis for our physics-informed loss.

    The \textbf{mass conservation loss} penalizes deviations from the divergence-free condition of the velocity field:
    \begin{equation}
    \mathcal{L}_{\text{mass}} = \left\| \nabla \cdot \mathbf{u} \right\|_2^2 = \left\| \frac{\partial u_x}{\partial x} + \frac{\partial u_y}{\partial y} \right\|_2^2.
    \end{equation}

    The \textbf{momentum conservation loss} measures the residual of the steady incompressible Navier--Stokes equations (per unit mass) by capturing convective transport, pressure gradients, and viscous diffusion. In our implementation, each term is normalized by its mean absolute value to improve numerical stability during training:
    \begin{equation}
    \begin{aligned}
    \mathcal{L}_{\text{mom}}
    &=
    \left\|
    \frac{(\mathbf{u}\!\cdot\!\nabla) u_x}{\sigma_c}
    \;+\;
    \frac{\partial p}{\partial x}\frac{1}{\sigma_p}
    \;-\;
    \frac{\mu}{\rho}\frac{\nabla^2 u_x}{\sigma_v}
    \;-\; g_x
    \right\|_2^2
    \\
    &\quad+\;
    \left\|
    \frac{(\mathbf{u}\!\cdot\!\nabla) u_y}{\sigma_c}
    \;+\;
    \frac{\partial p}{\partial y}\frac{1}{\sigma_p}
    \;-\;
    \frac{\mu}{\rho}\frac{\nabla^2 u_y}{\sigma_v}
    \;-\; g_y
    \right\|_2^2 ,
    \end{aligned}
    \end{equation}
where $\sigma_c$, $\sigma_p$, and $\sigma_v$ denote the mean absolute values of the convective, pressure, and viscous terms, respectively, used for normalization. Here, $\nabla^2 u_x$ and $\nabla^2 u_y$ are computed component-wise (vector Laplacian), and $\mathbf{g} = (g_x, g_y)$ denotes the body force per unit mass. In the present simulations, no external force field is applied, and thus $\mathbf{g} = \mathbf{0}$.

    These components are combined with the MSE loss to form the total training objective:
    \begin{equation}
    \mathcal{L}_{\text{total}} = \mathcal{L}_{\text{MSE}} + \lambda_{\text{phys}} \left( \lambda_{\text{mass}} \mathcal{L}_{\text{mass}} + \lambda_{\text{mom}} \mathcal{L}_{\text{mom}} \right),
    \end{equation}
    where $\lambda_{\text{mass}}$ and $\lambda_{\text{mom}} $ control the balance between the two physical terms, and $\lambda_{\text{phys}}$ regulates the overall contribution of physics-based supervision.

    These scalar weights play two key roles. First, they address potential scale mismatches between different loss components-particularly between gradients of pressure and velocity fields-by normalizing their influence on training. Second, they provide finer control over the optimization dynamics, enabling us to tune the strength of physical regularization in a principled manner.

    To further improve training stability and convergence, we adopt a \textbf{warm-up strategy} for $\lambda_{\text{phys}}$. At the beginning of training, physics-based losses are down-weighted to allow the model to first capture coarse patterns from data. The value of $\lambda_{\text{phys}}$ is then gradually increased to its target level over a fixed number of iterations. This prevents early-stage overfitting to noisy residuals and allows the model to integrate physical constraints more effectively as it learns.

    In summary, the hybrid loss formulation equips AeroDiT with both statistical learning capacity and inductive physical bias, enabling more accurate and physically consistent predictions.

    \vspace{1em}

    \subsection{Overall Training} \label{sec:overall}

    As illustrated in Figure~\ref{fig:main}, the training pipeline of AeroDiT involves three major components: the condition encoder, the diffusion transformer (DiT), and the function space decoder. The encoder and decoder are pretrained using reconstruction objectives, while the DiT is subsequently trained using the loss function described in Section~\ref{sec:loss}.

    \textbf{The condition encoder} processes airfoil geometry and inflow conditions into latent features. These features are then injected into the DiT either via concatenation or attention mechanisms to condition the generative process.

    \textbf{The DiT module} models the conditional distribution $p(\boldsymbol{x} \mid \boldsymbol{\Psi})$ by denoising noisy flow fields. At each diffusion timestep, the model predicts the added noise, supervised by both MSE and physics-informed losses.

    \textbf{The function space decoder} transforms the denoised latent features back into physical fields, such as pressure and velocity. It is responsible for translating abstract features into spatially coherent predictions.

    The complete training loop is summarized below.

    \begin{algorithm}[H]
    \caption{Training DiT Model}
    \KwIn{Arguments \texttt{args} (batch size, epochs, etc.), Trained VAE encoder-decoder.}
    \KwOut{Trained DiT model}

    \SetAlgoLined
    Initialize distributed training environment\;

    Initialize DiT model, EMA model, optimizer, and trained VAE models\;

    \For{epoch in $1, \dots, \texttt{args.epochs}$}{
        \For{each batch (condition, physical field) $(x, y)$ in DataLoader}{
        
            Encode $x$ and $y$ into the latent space using VAE models\;
            
            Sample random time steps $t$\;
            
            Compute $\mathcal{L}_{\text{total}}$\;
            
            Backpropagate and update DiT model parameters\;
            
        }
    }
    \end{algorithm}

    With the training procedures described above, the next section focuses on the experimental evaluation.

    \section{Experimental Evaluation}\label{Section:Experimental}
    To evaluate the performance of AeroDiT, we conducted extensive experiments using real-world datasets. The following sections detail the experimental setup, present a discussion of the results, and delve into specific case studies for a more in-depth analysis.  All the source codes to reproduce the results are available on \href{https://github.com/AI4SciFoundation/DiT4Science}{https://github.com/AI4SciFoundation/DiT4Science}.

    \subsection{Experiment Settings} 
    \subsubsection{Datasets and Data preparation} 
    In this study, we utilize three datasets and a test set, as described in \cite{thuerey2020deep}. These datasets are generated using RANS simulations to compute the velocity and pressure distributions of incompressible flows around airfoils. The simulations cover a wide range of Reynolds numbers (\(Re = [0.5, 5]\) million) and angles of attack (\(\pm 22.5^\circ\)), providing a diverse set of aerodynamic conditions. A total of 1505 unique airfoil shapes were sourced from the UIUC database and combined with randomly sampled freestream conditions to create the training inputs. The training and test sets are strictly foil-profile disjoint, and the 30 airfoils used for testing do not appear in the 1475 geometries used for training and validation.  The RANS solutions were computed using the Spalart-Allmaras (SA) turbulence model, a widely used one-equation model for industrial applications. Boundary conditions are directly embedded into the finite-volume formulation without ghost cells, and gradients and Laplacians are discretized using standard Gauss schemes that remain consistent near masked regions.The computations were carried out with OpenFOAM, an open-source CFD software.

    All simulation cases share constant freestream density \(\rho\) and dynamic viscosity \(\mu\), so variations in Reynolds number arise solely from differences in the freestream velocity \(U_\infty\) and characteristic chord length \(c\). Before training, the velocity and pressure fields are scaled using a characteristic velocity \(v_{\mathrm{norm}}\), computed from the maximum magnitude of the input velocity components in each sample. Specifically, velocity components are divided by \(v_{\mathrm{norm}}\) and pressure by \(v_{\mathrm{norm}}^2\). All channels are then normalized using fixed or dataset-specific maxima determined from the training data, and the same normalization constants are applied during inference to ensure consistency. This normalization strategy enforces a consistent non-dimensional representation of the flow fields, allowing the model to implicitly learn across different Reynolds numbers without explicit conditioning.

    The first dataset is a regular dataset (Regular), which contains 220,000 images with a data size of (6, 128, 128). The six channels in each image represent: the initial velocity along the $x$-axis (channel 1), the initial velocity along the $y$-axis (channel 2), the airfoil mask (channel 3), pressure (channel 4), velocity along the $x$-axis (channel 5), and velocity along the $y$-axis (channel 6). The primary task of our approach is to train the model to correctly predict the pressure, $x$-axis velocity, and $y$-axis velocity based on the first three channels. The second dataset is an augmented dataset (Shear), which also consists of 220,000 images. It is generated by shearing the wing shape along the center axis by $\pm$15 degrees to expand the shape space seen by the network, providing a richer variety of inputs. The final dataset is a small subset of the regular dataset (Subset), containing 6,000 images. This subset is introduced to better assess the model's generalization ability when trained with a limited amount of data.

    The test set comprises 90 images, which do not appear in the training data and remain unseen during all stages prior to the testing phase. The Reynolds number distribution in the test set spans a wide range, from [$2.04\times 10^6$ to $2.00\times 10^7$], encompassing 30 different airfoil types. All test cases lie within the same high-\(\mathrm{Re}\), fully turbulent regime as the training data and therefore obey the same similarity laws, resulting in nearly identical distributions of non-dimensional flow fields. All the datasets used in this study are publicly available \cite{TUM_dataset_m1470791,TUM_dataset_m1459172}.%at the following URLs:  

    \subsubsection{Configuration} 
    We employ the DiT (Scalable Diffusion Models with Transformers \cite{peebles2023scalable}) model as the backbone for AeroDiT and adopt the same encoder and decoder models as in \cite{ho2020denoising}. 
    We use three configurations: DiT-B/2, DiT-L/2, and DiT-XL/2. The suffix "/2" indicates that the model operates on a latent feature grid that is downsampled by a factor of 2 in spatial resolution, achieved through a patch size of $2 \times 2$ during tokenization.

    Initially, we train two separate VAE-based encoder-decoder models for 100 epochs:
    \begin{itemize}
        \item \textbf{Condition encoder-decoder}: Input is a tensor of shape (3, 128, 128), where the channels represent initial x-velocity, initial y-velocity, and the airfoil mask.
        \item \textbf{Function space encoder-decoder}: Input is a tensor of shape (3, 128, 128), with channels representing pressure, x-velocity, and y-velocity.
    \end{itemize}
    After training, the conditional encoder is retained as the encoder and the spatial decoder as the decoder. These components are then used to train the DiT model for 2,500,000 iterations with a global batch size of 4 for 150 epochs using the Adam optimizer with an initial learning rate of $1\times10^{-4}$. The training dataset contained 220{,}000 unique samples, and an exponential moving average (EMA) of the model parameters with a decay factor of 0.999 was maintained during training and used for final evaluation to improve stability and generalization.

    Following previous works, we evaluate model performance using the \textbf{mean relative $L_2$ error}, defined as:
    \begin{equation}
    \text{Relative } L_2 \text{ Error} = \frac{\|\hat{\boldsymbol{x}} - \boldsymbol{x}\|_2}{\|\boldsymbol{x}\|_2},
    \end{equation}
    where $\hat{\boldsymbol{x}}$ is the predicted field and $\boldsymbol{x}$ is the ground truth.

    We also report the \textbf{maximum absolute error} in selected case studies, defined as:
    \begin{equation}
    \text{Max Abs Error} = \max_i |\hat{x}_i - x_i|.
    \end{equation}

    We adopt the standard DDPM sampling procedure during inference, implemented via \texttt{p\_sample\_loop()}. Although our codebase also includes DDIM-based alternatives, we found that DDPM offers slightly better accuracy and more stable predictions in our fluid dynamics setting. Therefore, DDPM is used consistently throughout all experiments reported in this paper.

    For completeness, we list the architecture details of the DiT variants used in this work:
    \begin{itemize}
        \item \textbf{DiT-B/2}: 12 transformer blocks, 768 hidden size, 12 attention heads, patch size = 2 (i.e., $64 \times 64$ token map).
        \item \textbf{DiT-L/2}: 24 transformer blocks, 1024 hidden size, 16 attention heads, patch size = 2.
        \item \textbf{DiT-XL/2}: 28 transformer blocks, 1280 hidden size, 16 attention heads, patch size = 2.
    \end{itemize}
    The patch size of 2 means the input latent tensor of size $128 \times 128$ is divided into $64 \times 64$ non-overlapping patches, each of size $2 \times 2$. Each patch is then embedded and processed as a token within the DiT transformer backbone.

    \begin{table*}[ht]
        \centering
        \caption{Mean Relative $L_2$ Error Across Multiple Datasets.}
        \setlength\tabcolsep{2pt}
        \begin{tabular}{l|c c c|c c c|c c c}
            \hline
            &\multicolumn{3}{c|}{\textbf{DiT-B/2}}&\multicolumn{3}{c|}{\textbf{DiT-L/2}}&\multicolumn{3}{c}{\textbf{DiT-XL/2}}\\
            \hline
            Dataset&\textbf{p} & \textbf{ux} & \textbf{uy} &\textbf{p} & \textbf{ux} & \textbf{uy}&\textbf{p} & \textbf{u} & \textbf{uy}\\
            \hline
        Subset&$0.1927^{\pm0.23\%}$&$0.0262^{\pm0.02\%}$&$0.0865^{\pm0.03\%}$&$0.1989^{\pm0.42\%}$&$0.0285^{\pm0.14\%}$&$0.0909^{\pm0.33\%}$&$0.1935^{\pm0.13\%}$&$0.0271^{\pm0.07\%}$&$0.0884^{\pm0.05\%}$\\
            \hline
        Shear&$0.2000^{\pm0.39\%}$&$0.0470^{\pm0.22\%}$&$0.1120^{\pm0.30\%}$&$0.1964^{\pm0.19\%}$&$0.0458^{\pm0.13\%}$&$0.1116^{\pm0.08\%}$&$0.1864^{\pm0.76\%}$&$0.0415^{\pm0.24\%}$&$0.1050^{\pm0.37\%}$\\
            \hline
        Regular&$0.1065^{\pm0.58\%}$&$0.0251^{\pm0.22\%}$&$0.0547^{\pm0.22\%}$&$0.1041^{\pm0.54\%}$&$0.0250^{\pm0.23\%}$&$0.0537^{\pm0.34\%}$&$0.1103^{\pm0.26\%}$&$0.0278^{\pm0.22\%}$&$0.0578^{\pm0.31\%}$\\
            \hline
        \end{tabular}
        \label{tab:airfoil_reynolds2}
    \end{table*}

    \begin{figure*}[!t]
        \centering
        \includegraphics[width=1\linewidth]{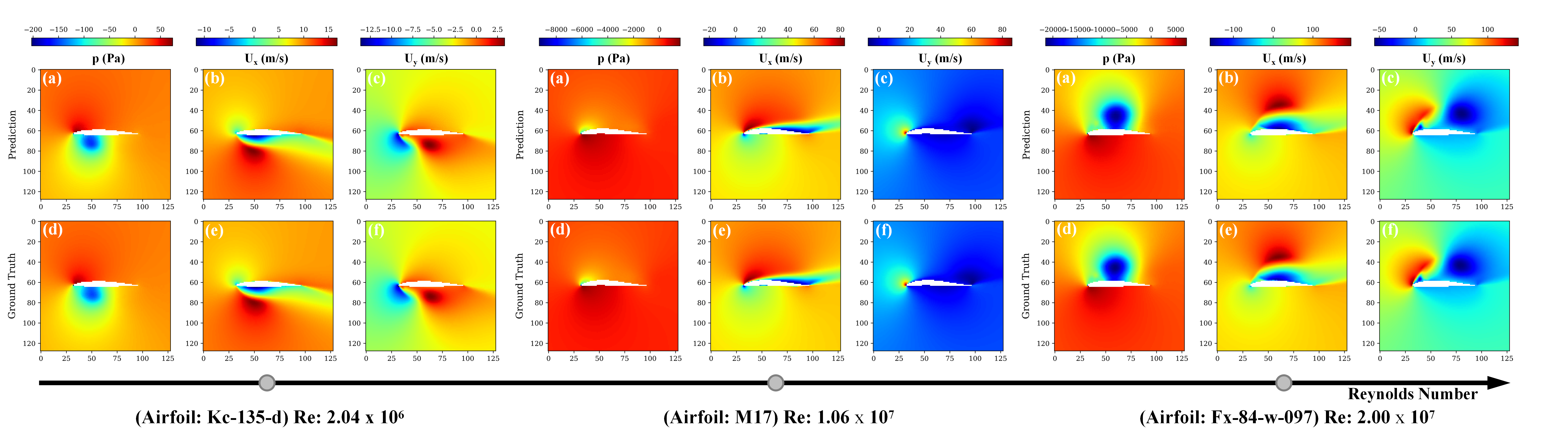}
        \caption{Visualization of input data (Airfoil geometry, initial horizontal velocity, initial parallel velocity), prediction (Pressure Field, velocity X, velocity Y) and the corresponding ground truth.}
        \label{fig:visual}
    \end{figure*}

    \subsection{Results and Discussion} 
    Table~\ref{tab:airfoil_reynolds2} summarizes the prediction results of the model using different DiT backbones. In experiments conducted on the annotated dataset, the best predicted $L_2$ errors were consistently below $0.105$, $0.025$, and $0.055$, respectively, across all test cases. Beyond $p, u_x, u_y$, we also report turbulence-relevant metrics (spanwise vorticity $\omega_z$, kinetic energy, enstrophy), computed via $3{\times}3$ centered differences over the fluid region; on the 90-case test set the mean relative $L_2$ errors are $0.0777$ ($\omega$), $0.0399$ (KE), and $0.0764$ (enstrophy).

    Figure~\ref{fig:visual} presents the performance of the DiT-L/2 model trained on the Regular dataset for three different airfoil configurations, corresponding to Reynolds numbers of \(2.04 \times 10^6\), \(1.06 \times 10^7\), and \(2.00 \times 10^7\), representing the minimum, median, and maximum Reynolds numbers in the test set. Additionally, the differences between the predicted and ground-truth values are visualized. The discrepancies between predictions and ground truth are minimal and nearly indistinguishable to the naked eye. Furthermore, the model's performance remains consistently reliable across both high and low Reynolds numbers, with negligible variation.

\begin{figure*}[!t]
        \centering
        \subfloat[Distributions of Reynolds number (top) and angle of attack (bottom) in training and test datasets]{
            \includegraphics[width=0.235\linewidth]{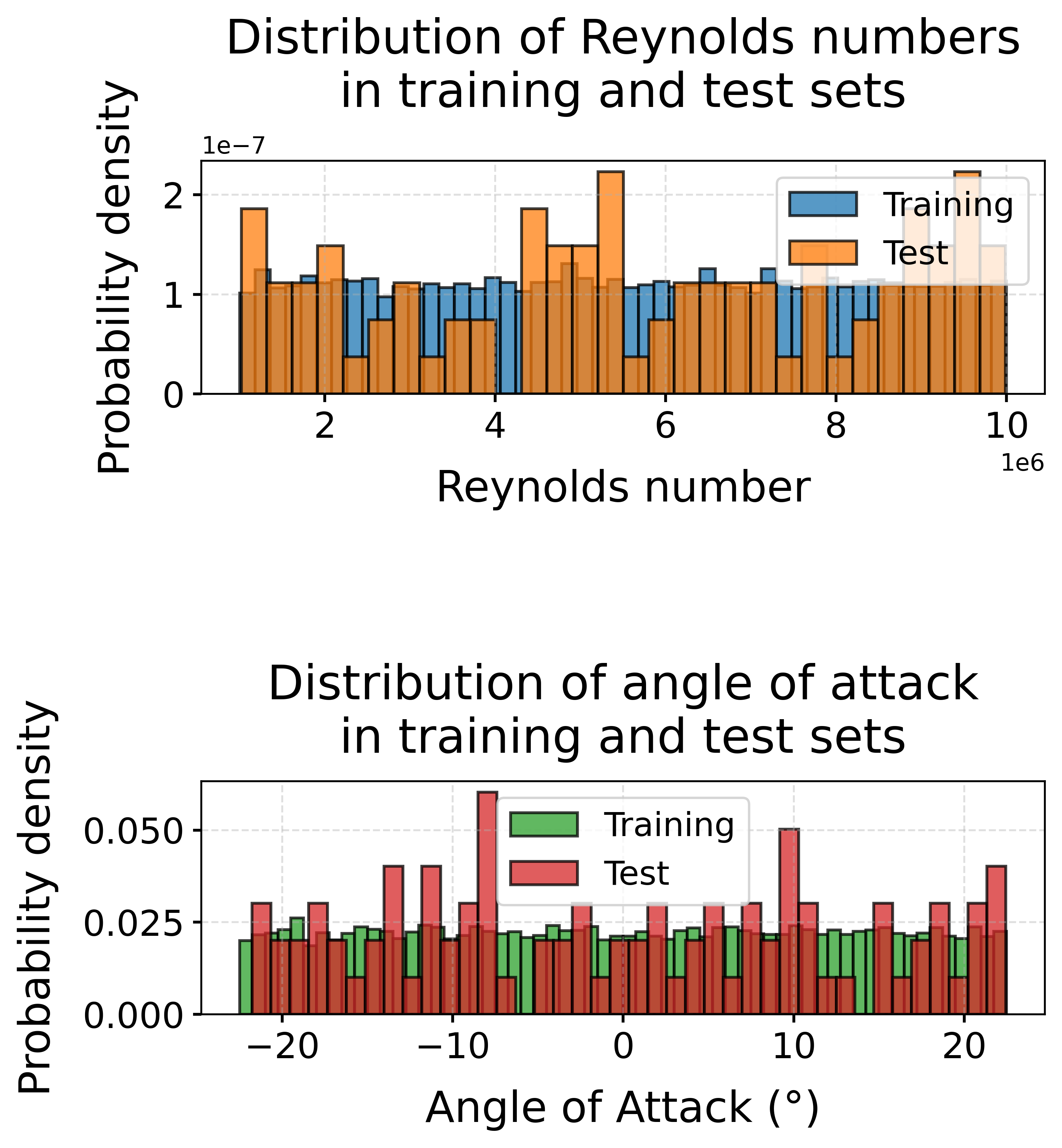}
            \label{fig:sub11}
        }
        \hfill
        \subfloat[Training loss convergence curve (DiT-L/2).]{
            \includegraphics[width=0.235\linewidth]{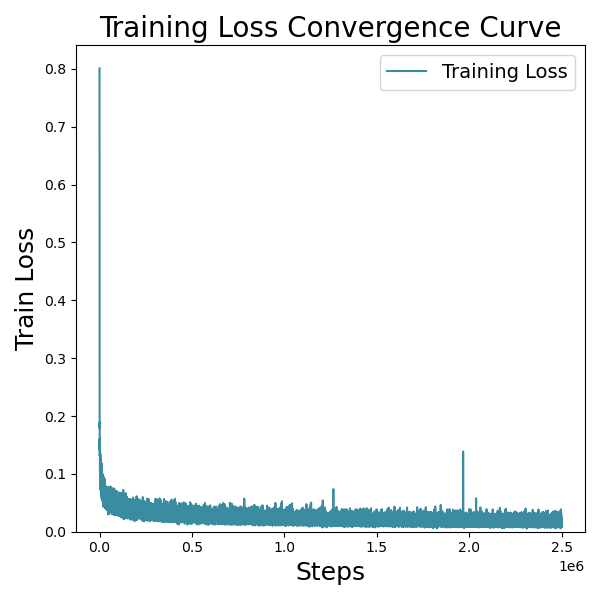}
            \label{fig:sub12}
        }\hfill	
        \subfloat[Error rate of $p, u_x, u_y$ w.r.t AeroDiT training steps (DiT L/2)]{
            \includegraphics[width=0.235\linewidth]{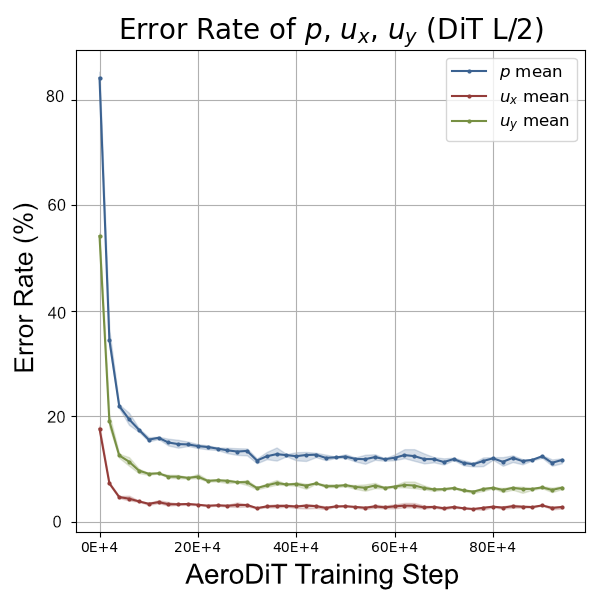}
            \label{fig:sub13}
        }	
        \hfill
        % \subfloat[$p, u_x, u_y$ and testing time w.r.t sampling steps (DiT-L/2)]{
        %     \includegraphics[width=0.235\linewidth]{fig/steps_.png}
        %     \label{fig:sub14}
        % }	
        \subfloat[Error rates of $p$, $u_x$, and $u_y$ and inference time as functions of AeroDiT diffusion sampling steps (DiT-L/2).]{
            \includegraphics[width=0.235\linewidth]{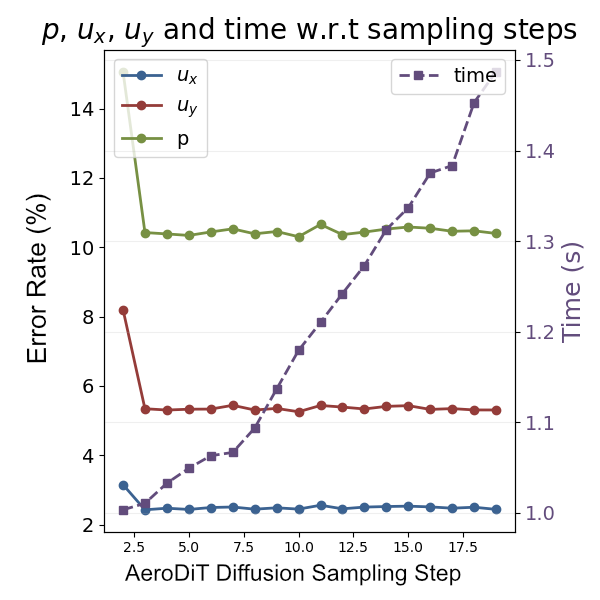}
            \label{fig:sub14}
        }

        \caption{Further studies on AeroDiT training and inference process.}
        \label{fig:V}
    \end{figure*}

    We provide more specific details on the model training process. Figure~\ref{fig:sub11} illustrates the distribution of Reynolds numbers and angles of attack in both training and test sets, confirming that all test cases lie within the global parameter range of the training data, with only a small fraction (\(\sim\)4--10\%) near the distribution boundaries. This indicates that the evaluation covers both in-distribution and mildly out-of-distribution scenarios, ensuring a robust assessment of generalization. Figure~\ref{fig:sub12} shows the variation of the experimental loss with the number of training iterations for the backbone network, revealing that the loss stabilizes after approximately \(5 \times 10^5\) iterations. Finally, Figure~\ref{fig:sub13} demonstrates that the model's performance on the test domain becomes stable after more than \(2 \times 10^5\) training iterations.

    Figure~\ref{fig:sub14} evaluates the effect of the number of diffusion sampling steps on the predictive performance of AeroDiT. The plot explicitly shows the error rates of pressure ($p$), streamwise velocity ($u_x$), and transverse velocity ($u_y$) as functions of the number of diffusion steps, along with the corresponding inference time. We observe that the prediction error decreases sharply within the first three steps and then plateaus, indicating diminishing returns with additional sampling. Beyond this point, accuracy improvements are marginal, while inference time continues to increase linearly. At around three to five steps, AeroDiT achieves near-optimal accuracy while generating predictions within a few seconds, enabling efficient near-real-time inference on the test set.

    \subsection{Computation Efficiency Analysis}
    All experiments were conducted on an NVIDIA RTX 6000 GPU. Tables~\ref{memory} and~\ref{tab:latency} together provide a comprehensive overview of the computational efficiency of AeroDiT. Table~\ref{memory} compares different DiT model variants (DiT-XL/2, DiT-L/2, DiT-B/2) in terms of GPU memory usage, total number of parameters, and training speed. As expected, smaller models (e.g., DiT-B/2) require significantly less GPU memory and achieve higher training throughput, while larger models (e.g., DiT-XL/2) provide greater representational capacity at a higher computational cost.
    
    Complementing these training-time metrics, Table~\ref{tab:latency} quantifies AeroDiT’s inference performance under different diffusion sampling steps. Even with increased sampling counts, inference remains efficient — requiring on the order of 1–2~s per sample and achieving throughputs of up to \(\sim 1\)~sample/s. For comparison, a traditional RANS simulation using OpenFOAM typically requires more than 70~s per case under similar conditions \cite{TUM_dataset_m1470791}. This highlights that AeroDiT not only scales efficiently during training but also delivers near-real-time inference performance suitable for rapid aerodynamic evaluation.

    \begin{table}[ht]
        \centering
        \caption{Latency and throughput of AeroDiT inference under different diffusion sampling steps (single RTX 6000 GPU).}
        \begin{tabular}{c|c|c}
            \hline
            \textbf{Sampling Steps} & \textbf{Latency (s)} & \textbf{Throughput (samples/s)} \\
            \hline
            5  & \(1.05 \pm 0.03\)  & \(0.95 \pm 0.03\) \\
            10 & \(1.20 \pm 0.03\)  & \(0.83 \pm 0.02\) \\
            20 & \(1.50 \pm 0.04\)  & \(0.66 \pm 0.02\) \\
            30 & \(1.80 \pm 0.05\)  & \(0.55 \pm 0.02\) \\
            50 & \(2.20 \pm 0.06\)  & \(0.45 \pm 0.01\) \\
            \hline
        \end{tabular}
        \label{tab:latency}
    \end{table}

    \begin{table}[h]
    \centering
    \caption{Training resource consumption.}
    \begin{tabular}{c|c|c|c}
    \hline
    \textbf{Model Type} &\textbf{Parameters} & \textbf{GPU Memory Usage} & \textbf{Training Steps/Sec} \\ \hline
    DiT-XL/2 &673,774,880& 21,265 MiB & 7.79 \\ \hline
    DiT-L/2  &456,898,592& 14,037 MiB & 10.89 \\ \hline
    DiT-B/2  &129,609,248& 5,535 MiB  & 23.67 \\ \hline
    \end{tabular}   
    \label{memory}
    \end{table}

    \subsection{Case Study}
To evaluate the performance of AeroDiT, we compare predictions against CFD benchmarks across various airfoil geometries and Reynolds numbers, as shown in Table~\ref{tab:airfoil_reynolds}. Note that the selected airfoils and Reynolds numbers belong to the target domain, distinct from the source domain used for training. This ensures that the evaluation accurately reflects the model's ability to generalize to unseen conditions. The chosen test cases assess generalization across diverse airfoil shapes and a wide range of Reynolds numbers (2.04--20.00 million), representative of the complex turbulent flows encountered in practical applications.

\begin{table}[ht]
    \centering
    \caption{Configuration of the airfoil and Reynolds number in the test cases.}
    \begin{tabular}{c|c|c}
        \hline
        \textbf{Test Case} & \textbf{Airfoil} & \textbf{Reynolds Number} \\
        \hline
        1 & Drela AG09 & \(4.04 \times 10^6\) \\
        2 & AH 63-K-127/24 & \(9.11 \times 10^6\) \\
        3 & EPPLER 59 & \(1.21 \times 10^7\) \\
        \hline
    \end{tabular}
    \label{tab:airfoil_reynolds}
\end{table}

\begin{figure}[!t]
    \centering
    \includegraphics[width=1\linewidth]{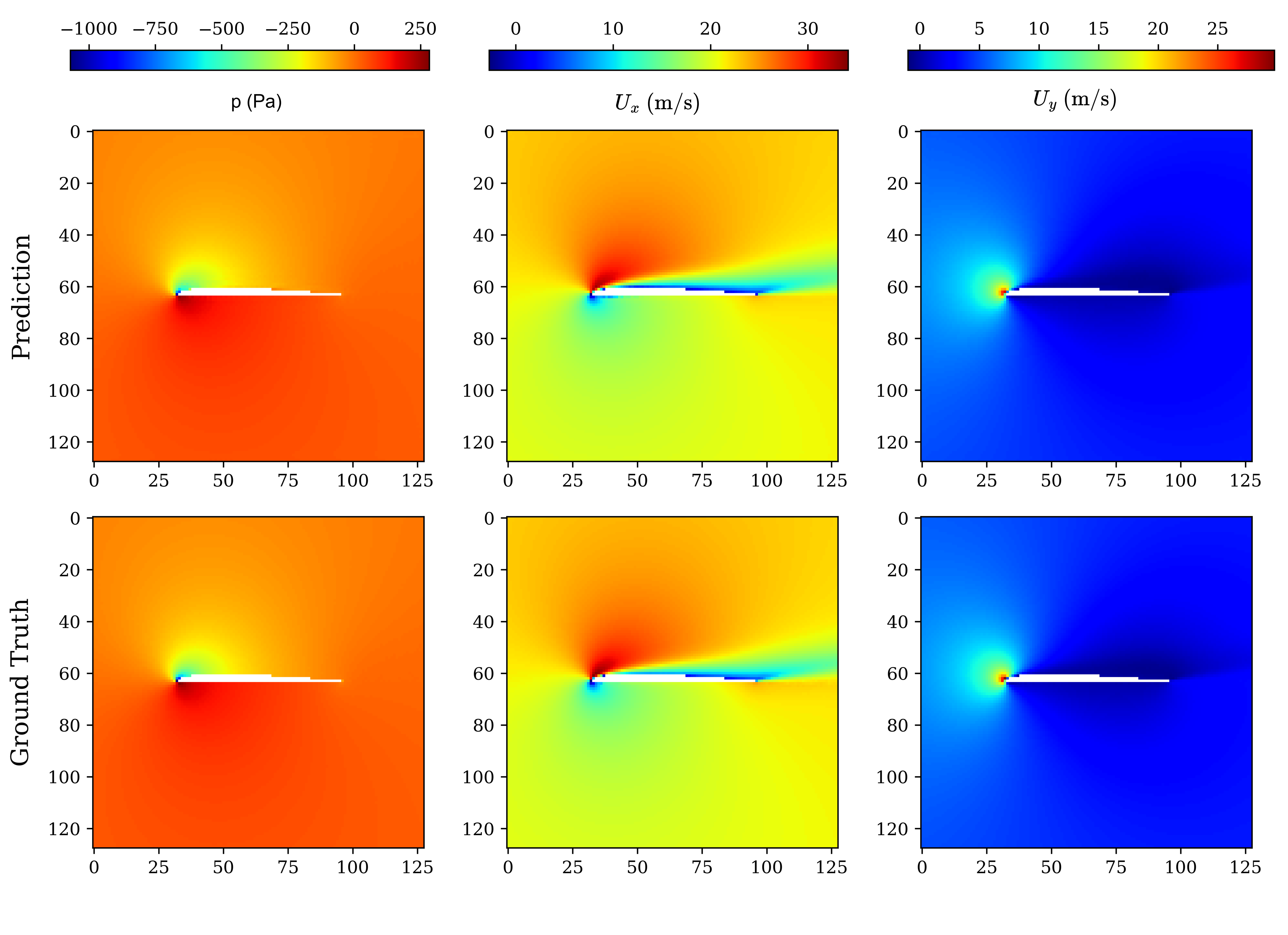}
    \caption{Normalized flow field of Drela AG09 airfoil at \(Re=4.04\times10^6\).}
    \label{fig:6}
\end{figure}

\begin{figure}[!t]
    \centering
    \includegraphics[width=1\linewidth]{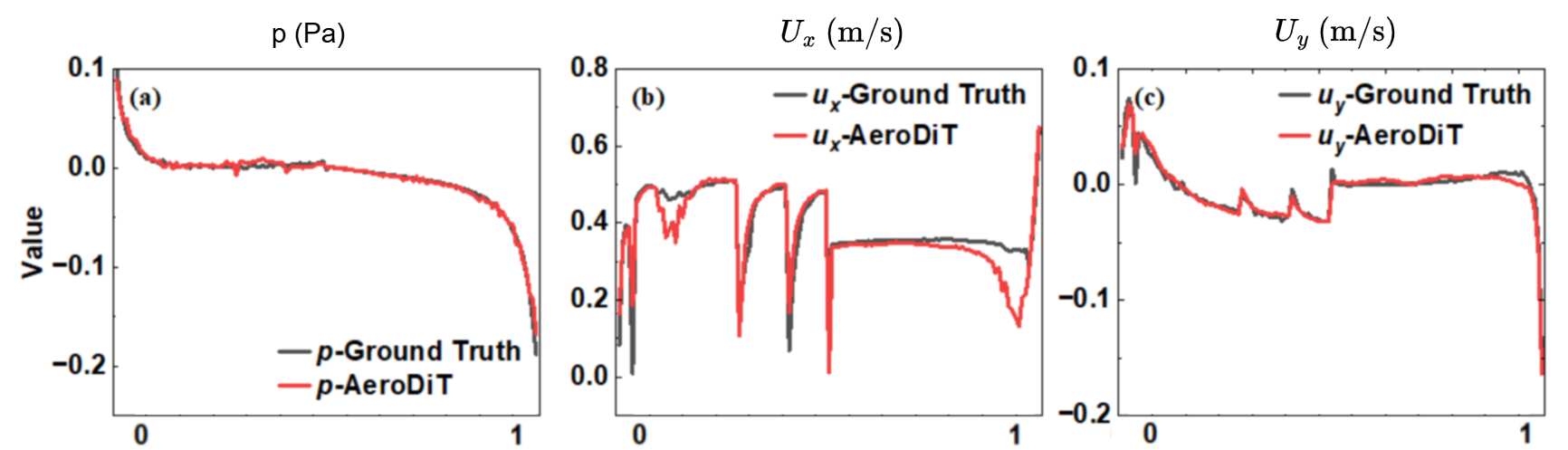}
    \caption{Comparison of surface pressure and velocities of Drela AG09 airfoil at \(Re=4.04\times10^6\). The horizontal axis represents the normalized arc-length coordinate (0–1) measured along the airfoil surface from the leading edge to the trailing edge.}
    \label{fig:7}
\end{figure}

Figure~\ref{fig:6} shows the normalized pressure and velocity fields predicted by AeroDiT and the CFD solution for the Drela AG09 airfoil at \(Re=4.04\times10^6\). Qualitatively, AeroDiT is in good agreement with CFD. For the pressure field, the maximum absolute error is around 0.09, while for the velocity field in both directions, the corresponding errors do not exceed 0.010 and 0.041. We observe that the largest errors concentrate along the \textit{airfoil surface}. To quantify these, Figure~\ref{fig:7} compares pressure and velocity along the surface: the streamwise velocity is captured best, whereas differences appear in surface pressure and transverse velocity. Physically, these residuals likely stem from boundary-layer behavior and potential local separation at high Reynolds numbers, where fine-scale turbulent structures dominate near-wall dynamics.

\begin{figure}[!t]
    \centering
    \includegraphics[width=1\linewidth]{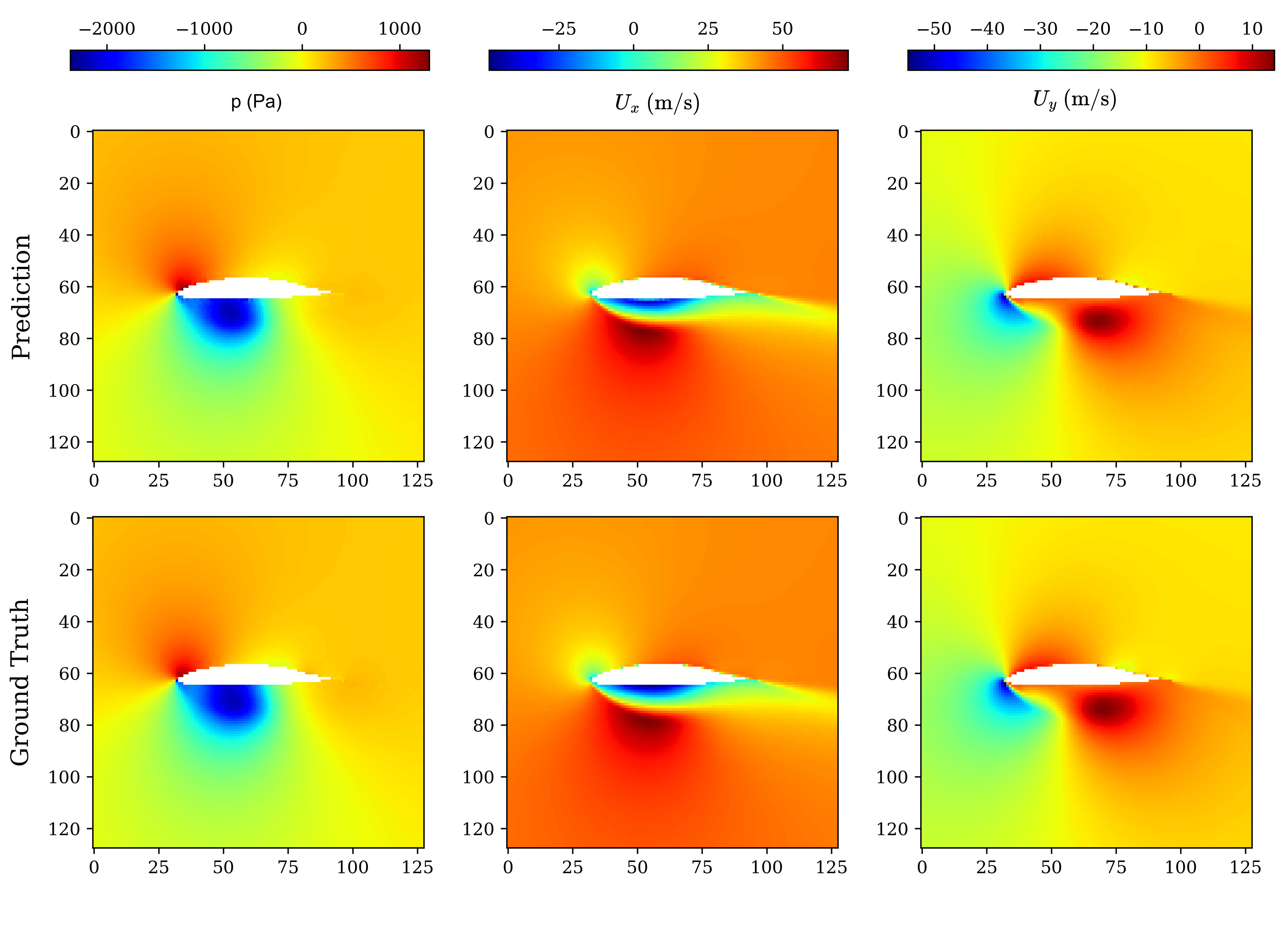}
    \caption{Normalized flow field of AH 63-K-127/24 airfoil at \(Re=9.11\times10^6\).}
    \label{fig:8}
\end{figure}

\begin{figure}[!t]
    \centering
    \includegraphics[width=1\linewidth]{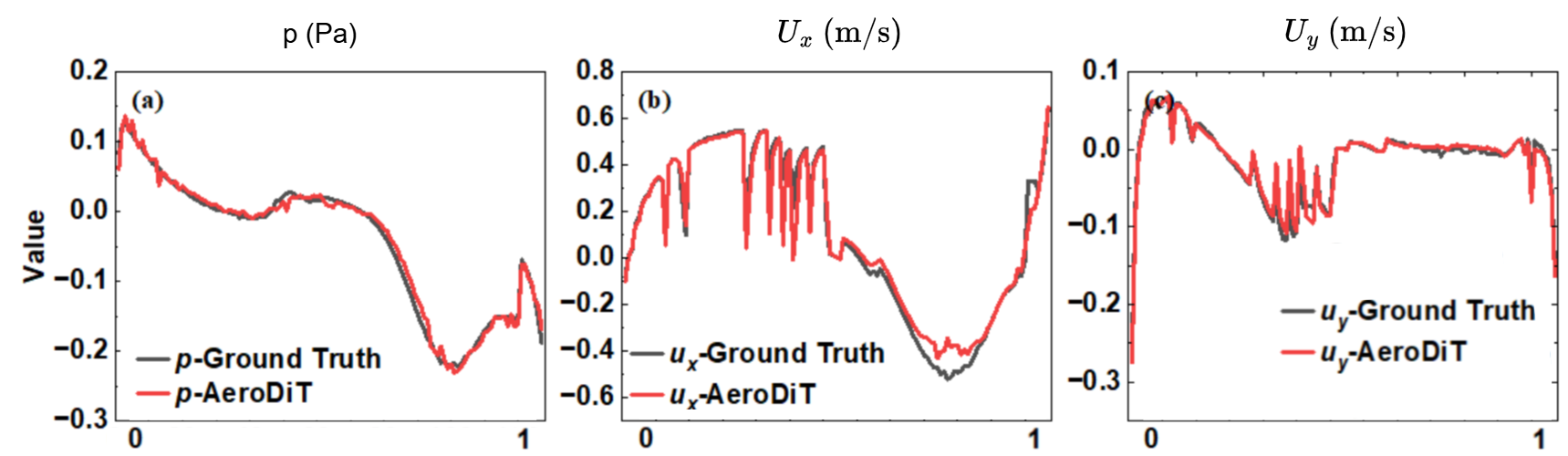}
    \caption{Comparison of surface pressure and velocities of AH 63-K-127/24 airfoil at \(Re=9.11\times10^6\). The surface coordinate is again shown as a normalized arc length (0–1) along the airfoil profile.}
    \label{fig:9}
\end{figure}

In test case 2, the airfoil is changed to AH 63-K-127/24 and the Reynolds number is increased to \(9.11\times10^6\). The normalized flow fields are shown in Figure~\ref{fig:8}. Somewhat counterintuitively, AeroDiT performs even better than in test case 1: absolute errors are mostly below 0.075 (pressure), 0.020 (\(u_x\)), and 0.045 (\(u_y\)). The surface comparisons in Figure~\ref{fig:9} confirm the improved curve match. A plausible explanation is that the training data distribution contains more samples in the vicinity of this Reynolds-number regime (order \(10^7\)), which improves conditional coverage and reduces extrapolation. From a physics perspective, the boundary layer remains attached over a larger portion of the chord in this case, mitigating near-wall discrepancies.

    \begin{figure}[!t]
        \centering
        \includegraphics[width=1\linewidth]{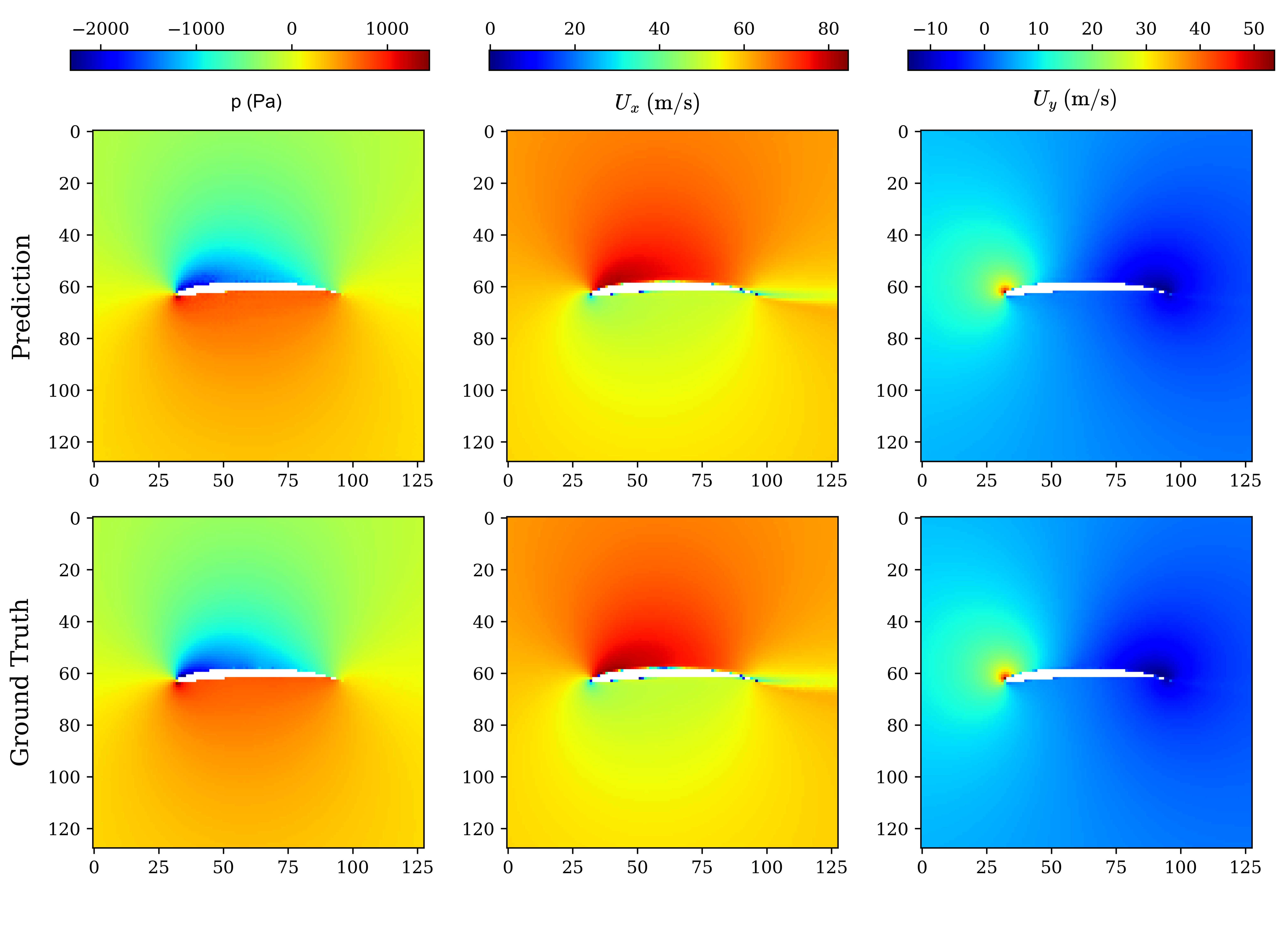}
        \caption{Normalized flow field of EPPLER 59 airfoil at \(Re=1.21\times10^7\).}
        \label{fig:10}
    \end{figure}

    \begin{figure}[!t]
        \centering
        \includegraphics[width=1\linewidth]{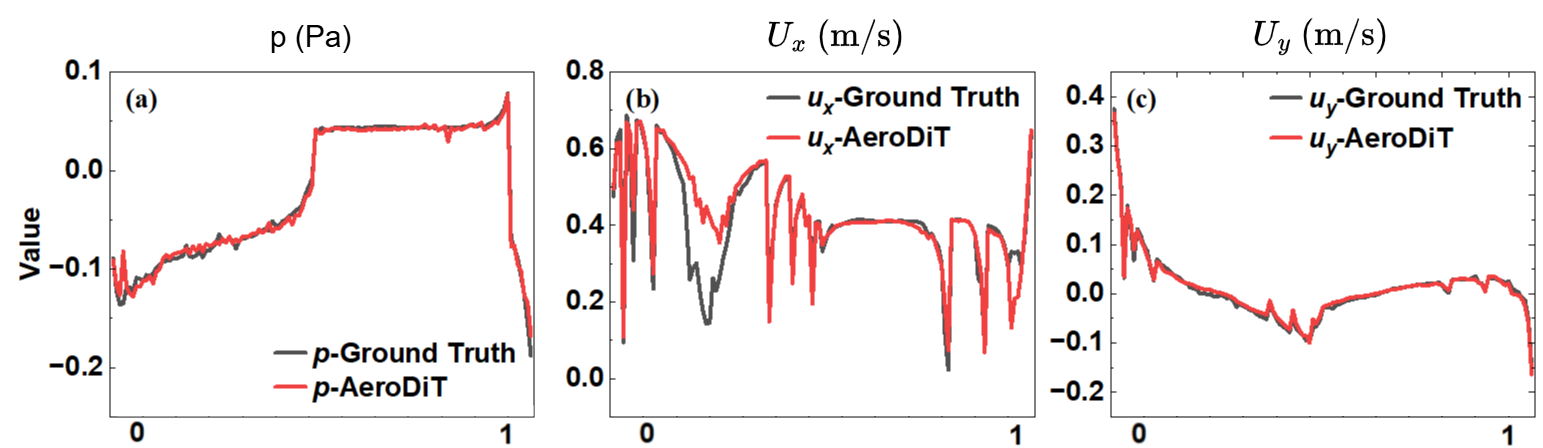}
        \caption{Comparison of surface pressure and velocities of EPPLER 59 airfoil at \(Re=1.21\times10^7\).The surface coordinate is expressed as a normalized arc-length parameter ranging from 0 to 1 along the airfoil profile.}
        \label{fig:11}
    \end{figure}

    Finally, the airfoil is EPPLER 59 at \(Re=1.21\times10^7\). The flow-field and surface quantity comparisons are shown in Figures~\ref{fig:10} and~\ref{fig:11}. Here the model performance degrades, especially in surface pressure and transverse velocity. Two factors likely contribute: (i) fewer high-\(Re\) cases at this order in the training set, and (ii) pronounced changes in flow physics when moving from \(O(10^6)\) to \(O(10^7)\), including stronger adverse-pressure-gradient effects and earlier separation onset. These observations suggest that adding more high-\(Re\) training samples and/or strengthening physics constraints (e.g., RANS-residual losses) could further improve robustness in this regime.

    Although $C_L$, $C_D$, and $C_m$ are not explicitly plotted, their values follow directly from the predicted surface pressure and shear distributions, which closely match CFD results. The accurate capture of pressure gradient reversals and near-wall flow features further ensures reliable identification of separation and reattachment, linking field-level accuracy to performance-critical quantities such as lift, drag, and pitching moment. While integral mass and momentum residuals were not explicitly computed, the close agreement in surface pressure and velocity distributions (Figs.~\ref{fig:7}, \ref{fig:9}, and~\ref{fig:11}) provides indirect evidence of global flow consistency around the airfoil surface. Future work will incorporate explicit global conservation constraints to enhance physical fidelity further.

\section{Discussion}\label{Section:Discussion}
This study demonstrates the significant potential of the Diffusion Transformer in improving airflow simulations for high Reynolds number airfoil flows. We evaluated AeroDiT against traditional CFD results using three distinct airfoil types and Reynolds numbers.

For the Drela AG09 airfoil at a Reynolds number of \( 4.04 \times 10^6 \), AeroDiT showed strong agreement with CFD, with a maximum absolute error of approximately 0.02 in the pressure field and no more than 0.06 in the velocity field. In a more challenging test case with the AH 63-K-127/24 airfoil at a Reynolds number of \( 9.11 \times 10^6 \), AeroDiT performed even better, with nearly all errors in the pressure and velocity fields remaining below 0.02. However, for the EPPLER 59 airfoil at a Reynolds number of \( 1.21 \times 10^7 \), performance was less satisfactory, likely due to the scarcity of high Reynolds number cases in the training dataset and the significant changes in flow field characteristics at this scale. In addition, we observe that part of the residual error is concentrated in the near-wall regions, which is expected given the current spatial resolution of \(128 \times 128\). This grid spacing imposes a limit on the model’s ability to resolve steep pressure gradients and velocity shear within the boundary layer. Furthermore, the latent space representation may introduce some smoothing of fine-scale wall features. While these effects do not significantly affect global aerodynamic predictions, they highlight the importance of resolution-aware modeling when targeting wall-resolved accuracy.

Beyond predictive accuracy, one notable advantage of AeroDiT lies in its inference efficiency. While diffusion transformers are indeed computationally intensive during training, our model only requires a single training run to generalize across diverse airfoil shapes and flow conditions. Once trained, AeroDiT generates physically consistent flow fields within seconds on a single GPU, which is orders of magnitude faster than traditional CFD solvers such as RANS. This enables rapid design evaluation and real-time inference in practical applications, demonstrating a substantial computational gain in engineering workflows.

U\mbox{-}Nets remain strong, widely used denoisers in diffusion models \cite{ho2020ddpm,rombach2022ldm}, whereas AeroDiT adopts a DiT backbone, which scales well and captures long\mbox{-}range, multi\mbox{-}scale dependencies \cite{peebles2023dit,pereira2024vitreview}. We therefore position AeroDiT as complementary to U\mbox{-}Net surrogates rather than a direct replacement.

Future directions could involve expanding the training dataset to include a broader range of high Reynolds number scenarios to enhance the model's robustness and accuracy in more complex flow fields. Future work will also explore higher spatial resolutions, multi-scale latent encoding, and wall-focused representations to further improve near-wall fidelity, as well as developing domain generalization techniques \cite{2023letting, wang2022generalizing} based on varying Reynolds numbers. By incorporating precise Reynolds numbers as conditional inputs, we aim to further improve the predictive performance of the framework.

\section{Conclusion}
This study introduces AeroDiT, a novel framework that applies Diffusion Transformers to make real-time, accurate flow field predictions. By leveraging DiTs' ability to learn conditional distributions \(p(\boldsymbol{x} | \boldsymbol{\Psi})\), AeroDiT achieves reliable predictions of pressure and velocity fields, with average \(L_2\) errors of 0.1 for pressure (\( p \)) and 0.03 and 0.05 for velocity components (\( u_x \), \( u_y \)), respectively. The results demonstrate the model's ability to generalize across a broad range of Reynolds numbers (\( 2.04 \times 10^6 \) to \( 2.00 \times 10^7 \)) and diverse airfoil geometries. In test cases involving the Drela AG09 and AH 63-K-127/24 airfoils, AeroDiT shows excellent agreement with CFD benchmarks, with prediction errors for pressure and velocity fields consistently below 0.02, even under challenging conditions.

Notably, AeroDiT exhibits strong performance in near-real-time scenarios, with the sampling steps converging to stable predictions within seconds. The transformer-based architecture significantly reduces computational costs, making high Reynolds number aerodynamic simulations more feasible for real-time applications. However, in cases involving extreme Reynolds numbers (\( > 1.2 \times 10^7 \)), the model's performance shows some limitations, likely due to the scarcity of corresponding training data and the drastic changes in flow field characteristics at these scales.

In addition, we incorporate explicit physics-informed loss terms based on RANS residuals, including mass and momentum conservation constraints. This integration improves the physical plausibility of predictions and enhances training stability, further supporting AeroDiT as a hybrid physics-ML modeling framework.

These findings highlight AeroDiT's potential to bridge the gap between computational efficiency and predictive accuracy in CFD simulations. Future work should focus on enriching the training dataset with higher Reynolds number scenarios and extending the model to unsteady and three-dimensional flow conditions, as well as exploring more complex physics constraints.

    \bibliography{aipsamp}% Produces the bibliography via BibTeX.

    \end{document}